\documentclass[aps,prc,twocolumn,showpacs,amsmath,amssymb,nofootinbib]{revtex4}
\usepackage {lscape}
\usepackage{graphicx}
\usepackage{amsmath}
\usepackage[export]{adjustbox}
\usepackage {color}
\usepackage{longtable}

\usepackage[bookmarks=true]{hyperref}
\usepackage{color}

  \hypersetup{
   colorlinks=true,
   pdfborder={0 0 0},
   citecolor=blue,
   linkcolor=blue,
   urlcolor=blue
 }

\begin{document}
\title{Shell model results for nuclear $\ensuremath{\beta^{-}}$ -- decay properties of $sd$ shell nuclei}

\author{Anil Kumar$^{1}$\footnote{anilkhichar102@gmail.com}, Praveen C. Srivastava$^{1}$\footnote{Corresponding author: pcsrifph@iitr.ac.in} and Toshio Suzuki$^{2,3}$\footnote{suzuki@chs.nihon-u.ac.jp}}

\address{$^{1}$Department of Physics, Indian Institute of Technology Roorkee,
Roorkee 247 667, India}

\address{$^{2}$Department of Physics, College of Humanities and Science, Nihon Univerity, Sakurajosui 3, Setagaya-ku, Tokyo 156-8550, Japan}

\address{$^{3}$National Astronomical Observatory of Japan, Osawa 2, Mitaka, Tokyo 181-8588, Japan}

\date{\hfill \today}

\begin{abstract}
  We evaluate the allowed $\beta^-$- decay properties of nuclei with $Z = 8 - 15$ systematically under the framework of the nuclear shell model
 with the use of the valence space Hamiltonians derived  from modern $ab~intio$ methods,
 such as in-medium similarity renormalization group and coupled-cluster theory.
  For comparison we also show results obtained with fitted interaction  derived from the chiral effective field theory and
phenomenological USDB interaction.
We have performed calculations for O $\rightarrow$ F, F $\rightarrow$ Ne, Ne $\rightarrow$ Na, Na $\rightarrow$ Mg, Mg $\rightarrow$ Al, 
Al $\rightarrow$ Si,
Si $\rightarrow$ P and P $\rightarrow$ S transitions.  
 Theoretical results of $B(GT)$, log$ft$ values and  half-lives,
are discussed and compared with the available experimental data.
\end{abstract}

\maketitle

\section{Introduction}


 Due to the recent progress in nuclear theory with the development of modern effective nucleon-nucleon interactions for $sd$ shell nuclei, it is now possible to predict nuclear observables with appropriate accuracy. 
There are recently many experimental data available for half-lives, log$ft$ values,  Gamow-Teller ($GT$) strengths,
$Q$ values and branching fractions \cite{nndc,ame2012,ame2016}. Thus it is highly desirable to
study $\beta^-$-decay properties using these newly developed interactions.
These theoretical and experimental developments also make it possible to evaluate quenching factor of the effective axial-vector coupling strength $g_A$
for single beta-decays.

The Gamow-Teller beta-decay transitions of $sd$-shell nuclei with five or more excess neutrons 
were predicted by Wildenthal et al in Ref. \cite{wcb}. 
More comprehensive study 
of $\beta$-decay properties of $sd$-shell nuclei for $A$ = 17--39 were reported by 
Brown and Wildenthal in Ref. \cite{brownwild}. 
In the middle of the $sd$ shell ($A = 28$) the effective matrix elements are quenched by an overall factor of 0.76$\pm$0.03, while 
the average quenching factor of  0.897$\pm$0.035 was obtained by Wilkinson \cite{wilkinson1} in a similar analysis for the mass region $A=6-21$.
The Gamow-Teller beta-decay rates for A$\leq$18 nuclei were reported in Ref. \cite{chou}.
In this work the effective Gamow-Teller operators are deduced for the $0p$ shell from a least-squares fit to 16 experimental matrix elements.
 In later years, the shell model calculations for $\beta$$^{-}$-decay properties 
of neutron-rich $Z=9-13$ nuclei with $N\geq 18$ were reported by Li and Ren in Ref. \cite{li}.
The importance of chiral two-body currents
in nuclei for the quenching of the Gamow-Teller
transitions and neutrinoless double-beta decay is
reported for a Fermi-gas model in Ref. \cite{Menendez}.
Theoretical calculations for half-lives of medium-mass
and heavy mass neutron-rich nuclei from QRPA
based on the Hartree-Fock Bogoliubov theory or
other global models are available in the literature
\cite{suhonen,q1,q2,q3,ser,marketin}.


The study of nuclei towards drip-line are of great interest and many studies on these nuclei have been observed. 
After more than twenty years exact location of drip-line for F and Ne recently confirmed in the RIKEN experiment \cite{ahn}.
In this region the ground state of several nuclei were recently confirmed, and more excited states were populated.
Study of `island of inversion' region attracted much attention from recent RIB-facility.  The intruder configuration 
is important for such nuclei for e.g. $^{28-31}$Ne, $^{30,31}$Na, $^{31-34,36}$Mg and $^{33}$Al isotopes. 
The beta decay half-lives of these nuclei become larger because of the influence of intruder configuration in these nuclei.
Due to strong deformation the wave functions of parent and daughter nuclei become different thus it reduces the $B(GT)$ values.



In the recent years ab initio approaches are most successful to predict nuclear structure properties of unstable nuclei. Thus it is worthwhile to
study $\beta$$^{-}$ -- decay properties using these ab initio approaches. 
In the present work, our aim is to study $\beta$$^{-}$ -- decay properties of $Z= 8-15$ nuclei 
corresponding to earlier and new experimental data within the framework of nuclear shell model using modern $ab~initio$  interactions.
The purpose of the present work is to study how well the recent $ab~initio$ and newly developed shell-model interactions based on chiral interactions can describe the $\beta$-decay properties in $sd$-shell, 
and also to find how much quenching is necessary for these interactions by comparing with many more experimental data than in Ref. \cite{archana_gt}.
 Effective values of $g_{A}$ for $sd$ model space corresponding to  $ab~initio$ interactions will be extracted.
This work will add more 
information to earlier works \cite{wcb,brownwild,wilkinson1,chou,li}, where shell model results with phenomenological effective interactions were reported.
Since the study of  $\beta$-decay properties based on $ab~initio$ methods is very limited, this is the first comprehensive study of 
 $\beta$$^{-}$ -- decay properties  in the $sd$-shell using $ab~initio$  interactions.
 The $ab~initio$ calculations of $GT$ strengths in $sd$ shell region for 13 different nuclear transitions including
electron-capture reaction rates for $^{23}$Na$(e^{-},\nu)$$^{23}$Ne and $^{25}$Mg$(e^{-},\nu)^{25}$Na were reported in Ref. \cite{archana_gt}. 
 
This paper is organized as follows. In Sec. II, we present details of $ab~initio$  interactions and the formalism for the $\beta$$^{-}$-decay properties.
In  Sec. III, we present theoretical results along with the experimental data. Finally,  a summary and conclusions are drawn in  Sec. IV.


\section{Theoretical Formalism}

\subsection{Ab~initio Hamiltonians}

To calculate $GT$, log$ft$ values and half-lives 
  for the $sd$ shell nuclei, we have performed shell-model 
calculations using two $ab~initio$ interactions : in-medium similarity renormalization group (IM-SRG)  \cite{stroberg,Tsukiyama}  and
coupled-cluster effective interactions (CCEI)  \cite{jansen,jan1}. Also   
we have performed calculations with newly fitted interaction  derived from the chiral effective field theory  \cite{Huth}.
For comparison,  we have also performed calculations with the phenomenological universal sd-shell Hamiltonian version B (USDB) effective interaction \cite{usdb}
in addition to  the above three interactions.
 For the diagonalization of matrices we used J-scheme shell-model code NuShellX\cite{nushellx}.
For the $ab~initio$ and USDB interactions, we have performed calculations in the $sd$ model space.

 The USDB starts from single-particle energies and two-body matrix elements, where the effects of three-nucleon interactions are considered to be included implicitly. 
The $ab~initio$ interaction, on the other hand, starts from chiral two-nucleon and three-nucleon interactions, and one-body and two-body terms outside a core are constructed.
The effects of the three-nucleon forces are thus more properly treated in the $ab~initio$ approach compared with the phenomenological one. 

Glazek and Wilson \cite{glazek} and Wegner \cite{wegner} developed techniques to diagonalize 
many-body Hamiltonians in free space known as the similarity renormalization group (SRG).
The SRG  consists of a continuous unitary transformation, parametrized by the flow parameter $s$,
and splits  the Hamiltonian $H(s)$   into diagonal and off-diagonal parts   
\begin{eqnarray}
H(s) = U^{\dagger}(s){H(0)}{U(s)} = H^{d}(s)+H^{od}(s) ,
\end{eqnarray}
where $H(s=0)$ is the initial Hamiltonian. Taking the derivative of the Hamiltonian with respect to $s$,  one gets 
\begin{eqnarray}
\frac{dH(s)}{ds}= [\eta(s),H(s)],
\end{eqnarray}
where
\begin{eqnarray}
\eta(s) = \frac{dU(s)}{ds}{U^{\dagger}(s)} = -\eta^{\dagger}(s),
\end{eqnarray}
is the anti-Hermitian generator of  the unitary transformation. For an appropriate value of $\eta(s)$, the off-diagonal part of the Hamiltonian, 
$H^{od}(s)$,  becomes zero as $s$ $\rightarrow\infty$. Instead of the free space evolution, in-medium SRG (IM-SRG) has  an attractive feature
that one can involve 3,...,A-body operators using only two-body mechanism. 
The starting Hamiltonian $H$ with respect to a finite-density reference state $|\Phi_{0}{\rangle}$ is given as

\begin{eqnarray}
H = E_0+\sum_{ij}{f_{ij}}\{{a_{i}^\dagger{a_j}}\}+\frac{1}{2!^2}\sum_{ijkl}{\Gamma_{ijkl}}\{{a_{i}^\dagger}{a_{j}^\dagger}{a_{l}}{a_{k}}\} \nonumber\\
+\frac{1}{3!^2}\sum_{ijklmn}{W_{ijklmn}}\{{a_{i}^\dagger}{a_{j}^\dagger}{a_{k}^\dagger}{a_{n}}{a_{m}}{a_{l}}\}.
\end{eqnarray}
Here the normal-ordered strings of creation and annihilation  
operators obey $\langle\Phi|\{a_{i}^{\dagger}{\cdots}a_{j}\}|\Phi\rangle$=0, 
and the $E_0$, $f_{ij},$ $\Gamma_{ijkl},$ and  $ W_{ijklmn} $ are the normal-ordered  zero-, one-, two-, and three-body terms, respectively
(see Ref. \cite{bogner,hergertpr,hergertps,lecture} for full details).
 In case of IM-SRG, targeted normal ordering with respect to the nearest closed shell rather than $^{16}$O is adopted to take into account the three-nucleon interaction among the valence nucleons.  

We  use another $ab$ $initio $ approach to study $\beta^{-}$-decay properties of nuclei in the $sd$ shell region, named as 
coupled-cluster effective interactions (CCEI). For this effective interaction, the intrinsic A-dependent Hamiltonian is given as (for IM-SRG interaction also): 
\begin{eqnarray}
\hat{H_{A}}=\sum_{i<j}\left(\frac{(\textbf{p}_{i}-\textbf{p}_{j})^2}{2mA}+\hat{V}_{\text{NN}}^{(i,j)}\right)+\sum_{i<j<k}\hat{V}_{\text{3N}}^{(i,j,k)}.
\end{eqnarray}

The $NN$ and $3N$ parts are taken from a next-to-next-to-next-to leading order (N3LO) chiral nucleon-nucleon interaction,
and a next-to-next-to leading order (N2LO) chiral three-body interaction, respectively.
For both IM-SRG and CCEI, we use $\Lambda_{\text{NN}}$ = 500 MeV for chiral N3LO $NN$ interaction \cite{entem,machleidt}, and 
$\Lambda_{\text{3N}}$ = 400 MeV for chiral N2LO $3N$ interaction \cite{navratil}, respectively. 

 In the CCEI to achieve faster model-space convergence,  the similarity renormalization group transformation has been used 
to evolve two-body and three-body  forces to the lower momentum scale $\lambda_{\text{SRG}} = 2.0$ $ \text{fm}^{-1}$ (see Ref. \cite{Jurgenson}
for further details). Also, for the coupled-cluster calculations, a Hartree-Fock basis built from thirteen major harmonic-oscillator orbitals
with frequency $\hbar\Omega$ = 20 MeV have been used.

We can expand the Hamiltonian for the suitable model-space using the valence-cluster expansion \cite{lisetskiy} given as
\begin{eqnarray}
H_{\text{CCEI}}^{\text{A}}= H_{0}^{A_{c}}+H_{1}^{A_{C}+1}+H_{2}^{A_{C}+2}+\cdots.
\end{eqnarray}
Here A is the mass  of the nucleus for which we are doing calculations, $H_{0}^{A_{C}}$ is the core  Hamiltonian,
$H_{1}^{A_{C}+1}$ is the valence one-body  Hamiltonian, and $H_{2}^{A_{C}+2} $ is 
the two-body Hamiltonian.
The two-body term is derived from Eq. (6) by using the Okubo-Lee-Suzuki (OLS) similarity transformation \cite{okubo,suzuki}. 
After using this unitary transformation the effective Hamiltonian become non-Hermitian. 

For changing the non-Hermitian to Hermitian effective Hamiltonian  the 
metric operator [$S^{\dagger}S$] = $P_{2}(1+\omega^{\dagger}\omega)P_{2}$ is used, where S is a matrix that diagonalize the Hamiltonians
(see Ref. \cite{nbarrett} for further details ).
After using the
metric operator the Hermitian shell-model Hamiltonian is then obtained as $[S^{\dagger}S]^{1/2}\hat{H}_{\text{CCEI}}^{A}[S^{\dagger}S]^{-1/2}$. 
  Using IM-SRG targeted for a particular nucleus \cite{imsrg_ragnar} and CCEI interactions, the shell model results
for spectroscopic factors and electromagnetic properties are reported
in Refs. \cite{pcs_prc,archana_prc}.
 In case of CCEI, the core is fixed to be $^{16}$O and no target normal ordering is carried out.


Recently, L. Huth et al. \cite{Huth} derived  a shell-model  interaction from chiral effective field theory.
The valence-space Hamiltonian for $sd$ shell is constructed as a general operators 
having two low energy constants (LECs) at leading order (LO) and seven new LECs at next-to-leading order (NLO)  and
fitted the LECs of CEFT operators directly to 441 ground- and excited-state energies. 
 For the chiral EFT interaction they have taken the expansion
 in terms of power of $(Q/\Lambda_{b})^{\nu}$ based on Weinberg power counting \cite{wibnernpb}, where
Q is a low-momentum scale or pion mass $m_\pi$ and $\Lambda_b$ $\sim$ 500 MeV is the chiral-symmetry-breaking scale.

\subsection{Beta-Decay Theory}

In the beta decay, the $ft$ value corresponding to $GT$ transition from the initial state $i$ of the parent nucleus to the final state $f$
in the daughter nucleus is expressed as  \cite{Piechaczek}

\begin{equation}
  f_{A}t_{i\rightarrow{f}} = \frac{6177}{[ B(GT;i\rightarrow{f})]},
\end{equation}
where $B(GT)$ is the Gamow-Teller  transition strength, and  $f_A$ is  the axial vector phase space
factor that  contains the lepton kinematics.  In this work, we have calculated the phase space factor $f_{A}$ with  parameters given by
Wilkinson and Macefield \cite{wilkinson} together with the correction factors given in Refs. \cite{Sirlin,wga}. The $ft$ values are very large, so they are
 defined in term of ``log$ft$" values. The log$ft$ is expressed as log$ft$= log$_{10}({f_{A}t_{i\rightarrow{f}}})$.

The total half-life $T_{1/2}$  is related to  the partial half-life as 
\begin{equation}
 \frac{1}{T_{1/2}}= {\sum_f {\frac{1}{t_{i\rightarrow{f}}}}},
\end{equation}
 where $f$ runs over all the possible daughter states that are populated through $GT$ transitions.

The partial half-life is related to the total half-life $T_{1/2}$ of the allowed $\beta^{-}$-decay as 
\begin{equation}
 t_{i\rightarrow{f}}= {\frac{T_{1/2}}{b_{r}}},
\end{equation}
where,  ${b_{r}}$ is called the branching ratio for the transition with partial half-life ${t_{i\rightarrow{f}}}$.

The Gamow-Teller strength $B(GT)$ is calculated using the following expression:
\begin{equation}
  B(GT;i\rightarrow{f})= (g_{A}^{\text{eff}})^2\frac{1}{2J_i + 1}  |{\langle {f}|| \sum_{k}{{\sigma}^k{\tau}_{\pm}^k} ||i \rangle}|^2. 
\end{equation}
 The effective axial-vector coupling constant is calculated from $g_{A}^{\text{ eff}} = q g_{A}^{\text{free}}$, where  $g_{A}^{\text{ free}}$ = -1.260, and $q$ is the quenching factor.
 The $|i\rangle$ and $|f\rangle$ are the initial and final state 
shell-model wave functions, respectively, here the $\tau_{\pm}$ refers to
isospin operator for the $\beta^{\pm}$ decay, 
for the $\beta^-$-decay we  use the convention  $\tau_-|n\rangle$ = $|p\rangle$,  $J_i$ is the initial-state angular momentum.


Following Refs.  \cite{brownwild,chou,pinedo}, we define

\begin{equation}
M(GT)= [(2J_{i}+1)B(GT)]^{1/2},
\end{equation}
which is independent of the direction of the transitions. 


 $R(GT)$ values are defined as

\begin{equation}
R(GT) = M(GT)/W,
\end{equation}
where the total strength $W$ is defined by 

\begin{equation}
  W=\left\{
  \begin{array}{@{}ll@{}}
    |g_{A}/g_{V}|[(2J_{i}+1)3|N_{i}-Z_{i}|]^{1/2} , & for N_{i} \neq  Z_{i},\\
    |g_{A}/g_{V}|[(2J_{f}+1)3|N_{f}-Z_{f}|]^{1/2} , & for N_{i} = Z_{i}.
  \end{array}\right.
\end{equation}

 Here $g_{V}$(=1.00) is the vector coupling constant; $N_{i}$ ($N_{f}$), $Z_{i}$ ($Z_{f}$) are 
neutron and proton number of initial (final) states, respectively and 
 $J_i$  $(J_f)$ is the angular momentum of the initial (final) state.
In the $\beta^{-}$-decay the endpoint energy of the electron  $E_0$ (in units of MeV) is  an essential quantity to calculate the phase space factor
$f_A$.
 $E_{0}$ is given by the expression \cite{brownwild}
\begin{equation}
 E_0=(Q+E_i)-E_f,
\end{equation}
 where the Q is the $\beta$-decay Q value, and $E_i $ and $E_f$ are  excitation energies of the initial and final states. 
Here, we have taken Q values from the experimental data \cite{ame2016}.

\begin{figure*}
\includegraphics[width=8.5cm,height=6.8cm,clip]{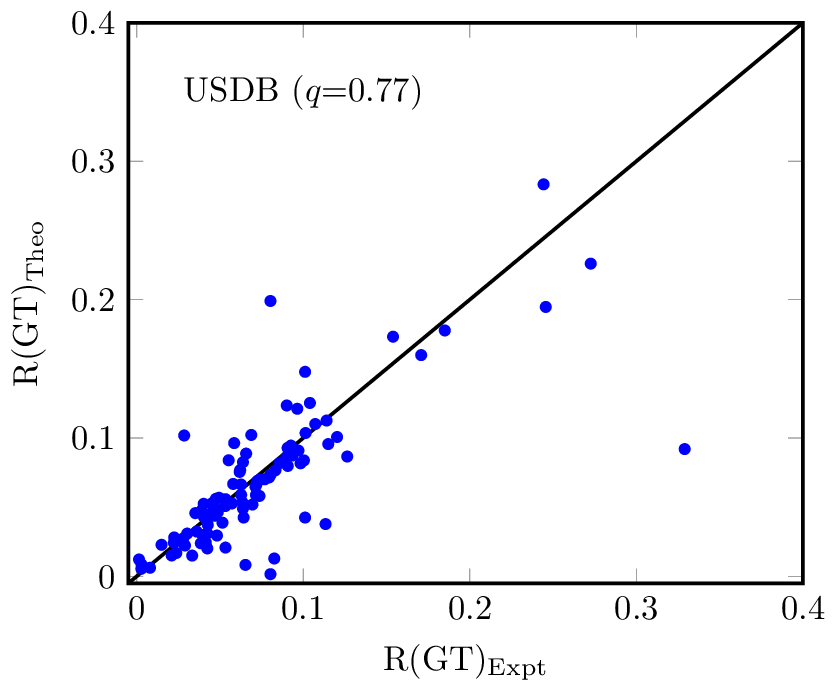}
\includegraphics[width=8.5cm,height=6.8cm,clip]{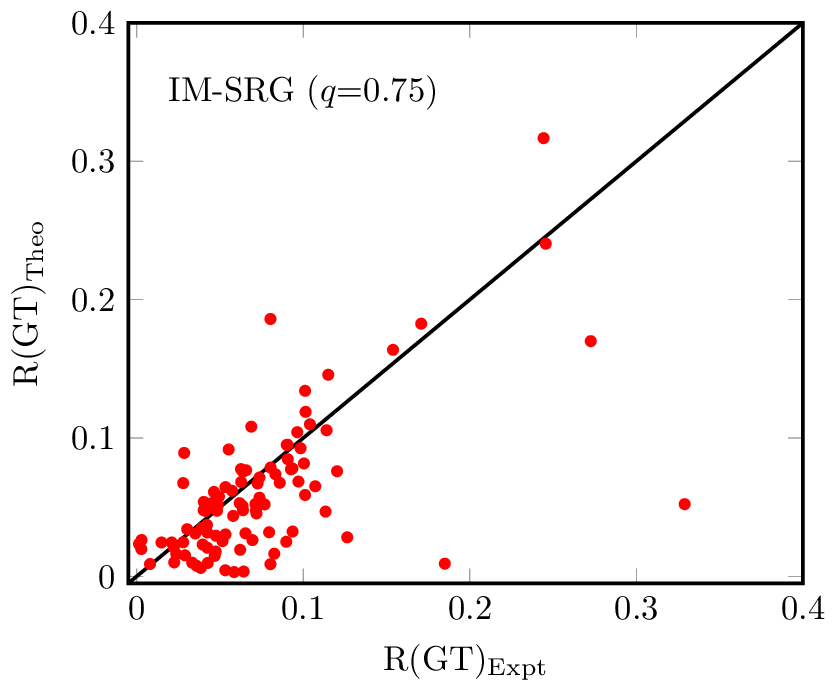}
\includegraphics[width=8.5cm,height=6.8cm,clip]{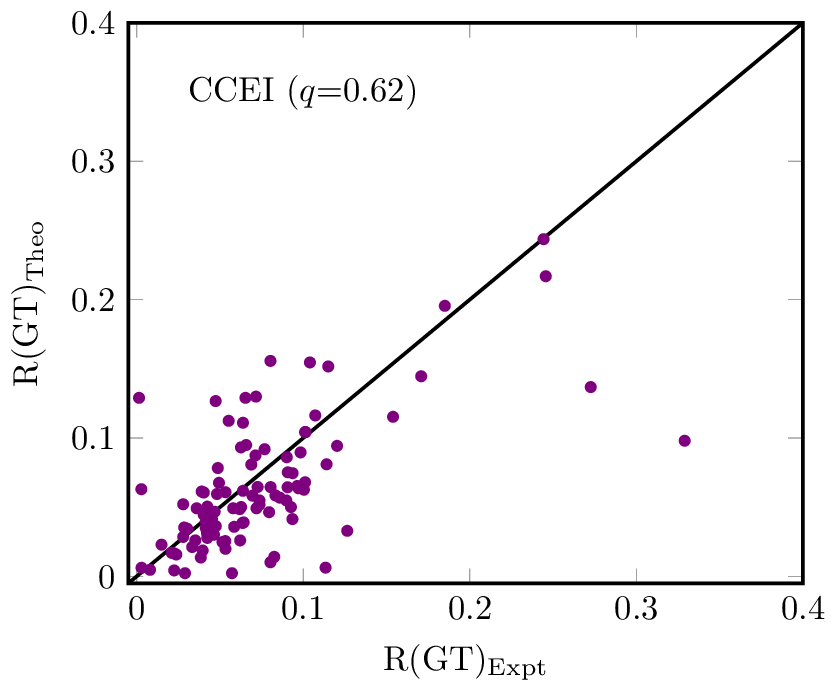}
\includegraphics[width=8.5cm,height=6.8cm,clip]{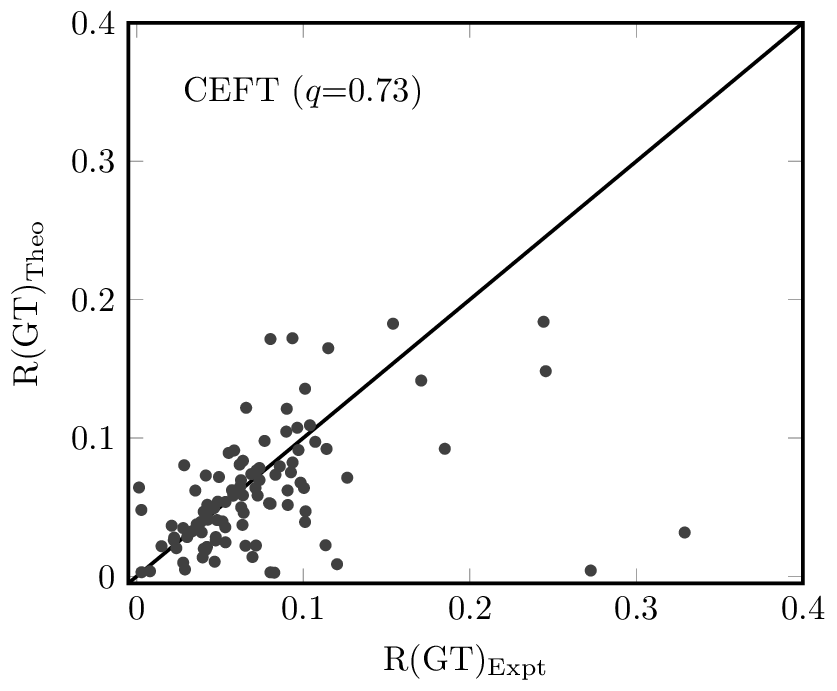}

 \caption{\label{qf_fig} (Color online) 
Comparison of the experimental values of the matrix elements $R(GT)$ with the theoretical ones obtained for the different effective interactions.
Each transition is indicated by a point. 
Experimental and theoretical  values are given by the horizontal and vertical coordinates, respectively. 
 When calculated and experimental R(GT ) values are the same, these points will be on the diagonal line.}
\end{figure*}

\section{Results and Discussions}
In Table~\ref{mgt} we compare calculated and experimental values of the matrix elements $M(GT)$.  Calculated values of $M(GT)$ presented here are those with $q$=1. 
 The $\beta$-decay energies ($Q$), branching  ratios ($I_{\beta}$) and log$ft$ values as well as the values of W are given in Table~\ref{mgt}.
The quenching factors are obtained by chi-squared fitting of the theoretical $R(GT)$ values to the corresponding experimental $R(GT)$ values. 
The quenching factors as well as the root-mean-square (RMS) deviations for the effective interactions considered here are given in Table~\ref{qf_tab}. 
The quenching factors  slightly change for different effective interactions.
 Their values are in the range of $q$ =0.62-0.77.
 The value for USDB, $q$=0.77$\pm$0.02, is consistent with the one, $q$ =0.764,  obtained for the USDB by shell model in te $sd$-shell as reported in Ref. \cite{Richter}.
 We have obtained the  RMS deviation values 0.0469, 0.0440, 0.0541 and 0.0356 corresponding to IM-SRG, CCEI, CEFT and USDB interactions, respectively.
We see that the RMS deviations for the $ab~initio$ and CEFT interactions are slightly higher than that for the USDB interaction.
 As a result, points of USDB in Fig.\ref{qf_fig} concentrate on the diagonal line.

 We have plotted the experimental $R(GT)$ values with respect to
the theoretical $R(GT)$ values for the $sd$ shell nuclei in Fig.\ref{qf_fig}. 
 In this figure there are some isolated points, e.g. around $R(GT)_\text{Expt}$ = 0.3.  This is because experimental $M(GT)$ value (expt. value of log$ft$ $\geq$ 3.3)
corresponding to transition $^{34}$Si$(0^+)$$\rightarrow$$^{34}$P$(1^+)$ is large as compared to theoretical value. This deviation is due to experimental uncertainty.
When calculated and experimental $R(GT)$ values are the same, then these points will be on the diagonal line.
 The different sources of renormalization \cite{suhonen_nndof,suhonen_nndof1,coraggio,Arima,Towner} affecting the values of $g_{A}$ depends on 
(i) missing configurations outside the $sd$-shell,
(ii) non-nucleonic degree's of freedom such as $\Delta_{33}$ resonance, and
 (iii) many-body operators induced by unitary transformations in the $ab~initio$ method.
  In the shell model we need effective value of $g_A$  to reproduce experimental results.
In our calculation corresponding to $|g_A^{\text{free}}|=1.26$,  we have obtained the value of
$|g_A^{\text{eff}}|$ as 0.97, 0.95, and 0.92 for USDB, IM-SRG, and CEFT interactions, respectively.
These values are close to unity, however, 
the CCEI interaction gives 0.78 far from the unity. 
For further calculations, we have used the effective value of $g_A$ corresponding to different interactions.

\begin{figure*}
\includegraphics[width=8.5cm,height=7.5cm,clip]{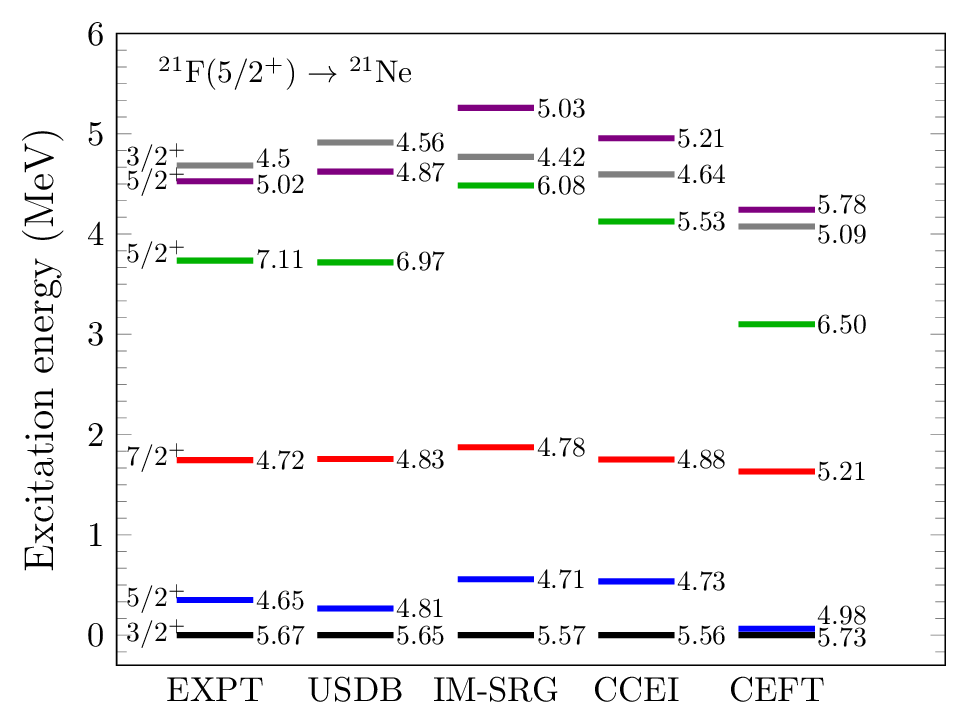}
\includegraphics[width=8.5cm,height=7.5cm,clip]{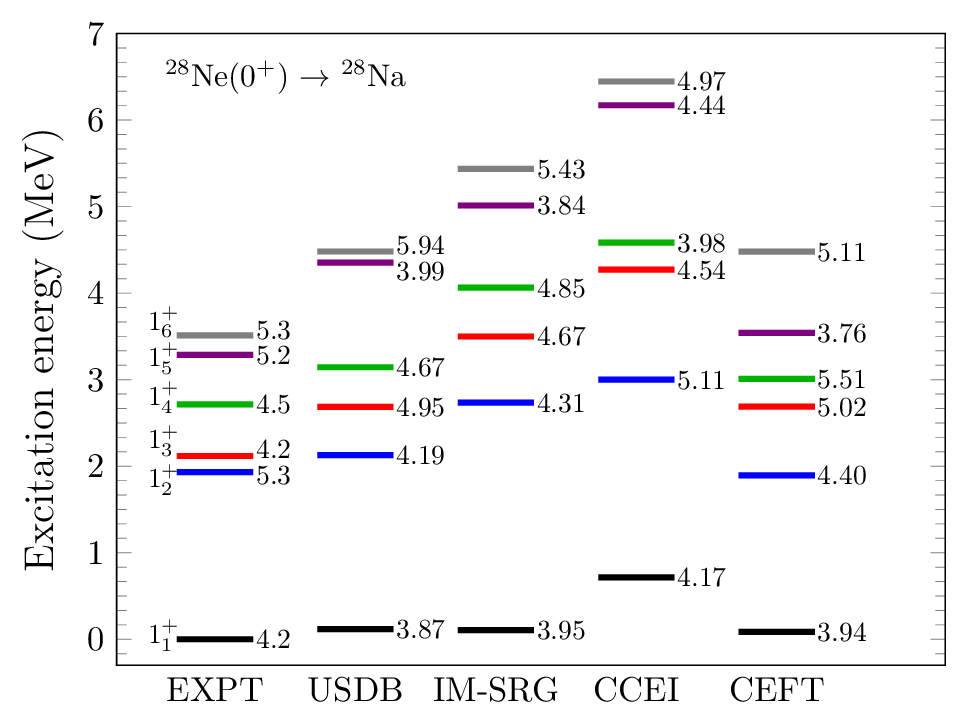}
\includegraphics[width=8.5cm,height=7.5cm,clip]{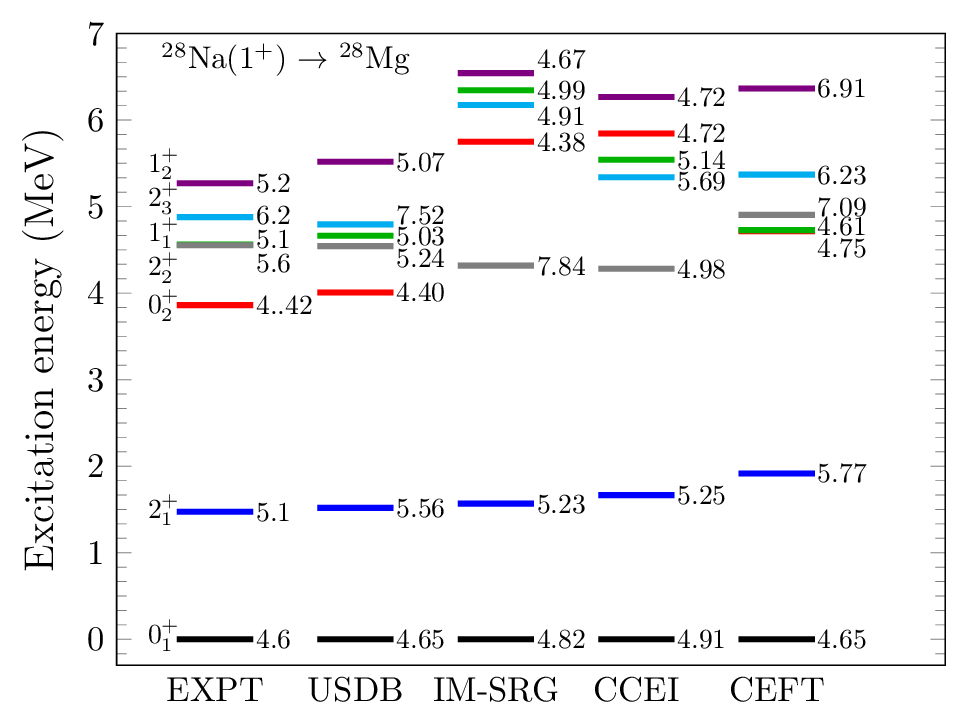} 
\caption{\label{logft} (Color online) 
Comparison experimental and theoretical (with different interactions)
distribution of log$ft$ values for the  $\beta^{-}$--decay for $^{21}$F, $^{28}$Ne and $^{28}$Na.}
\end{figure*}

\begin{table*}
\caption{\label{mgt} Experimental and theoretical $M$(GT) matrix elements. $I_\beta$ are the branching ratios.  
J$_n^\pi$ and T$_n^\pi$ are the spin-parity and isospin of the final states, respectively,  where n distinguishes the states with the same J in order of energy.
All other quantities are explained in the text.  The experimental data have been taken from \cite{nndc}.}
\begin{ruledtabular}
\begin{tabular}{ccccccccccccc}
 &                          & &  & & & & &  M(GT)    &    \\
\cline{7-11}

$^{A}$Z$_{i}(J^{\pi})$  & $^{A}$Z$_{f}$ & 2J$_n^\pi$,2T$_n^\pi$ &  Q (MeV) & $I_\beta${(\%)}  &log$ft$(exp.) & EXPT. & USDB & IM-SRG  & CCEI   &  CEFT   & W \\
\hline

$^{19}$O($5/2^+$) & $^{19}$F  & 7$^+$,1   & 0.442 & 0.0984(30) & 3.86(17) & 2.262 & 3.406 & 3.910 & 3.640 & 2.334 & 9.259 \\ 
&                             & 5$^+$,1   & 4.622 & 45.4(15) & 5.38(15) & 0.393 & 0.243 & 0.256 & 0.416 & 0.593 & \\
&                             & 3$_1^+$,1 & 3.266 & 54.4(12) & 4.62(10) & 0.939 & 1.245 & 1.468 & 1.556 & 0.596 & \\

$^{20}$O($0^+$)   & $^{20}$F  & 2$_1^+$,2 & 2.757 & 99.97(3) & 3.73(6) & 1.072 & 1.104 & 1.399 & 1.527 & 0.887 & 4.365 \\
&                             & 2$_2^+$,2 & 0.325 & 0.027(3)  & 3.64(6) & 1.190 & 1.281 & 0.989 & 0.963 & 0.025 &  \\

$^{21}$O($5/2^+$) & $^{21}$F  & 3$^+$,3   & 6.380 & 37.2(12)  & 5.22(2) & 0.473 & 0.464 & 0.367 & 0.360 & 0.223 & 11.953 \\

$^{22}$O($0^+$) & $^{22}$F    & 2$_1^+$,4 & 4.860 & 31(5)   & 4.6(1)  & 0.394 & 0.403 & 0.509 & 0.473 & 0.573 & 5.346 \\
&                             & 2$_2^+$,4 & 3.920 & 68(8)   & 3.8(1)  & 0.989 & 1.234 & 0.066 & 1.686 & 0.675 &  \\

$^{24}$O($0^+$) & $^{24}$F    & 2$^+$,6   & 9.700 & 40(4)  & 4.3(1)  & 0.556 & 0.990 & 0.782 & 0.857 & 1.024 & 6.173 \\

$^{20}$F($2^+$) & $^{20}$Ne   & 4$^+$,0   & 5.390 & 99.99(8)  & 4.97(11) & 0.575 & 0.687 & 0.679 & 0.651 & 0.694 & 6.901 \\

$^{21}$F($5/2^+$) & $^{21}$Ne & 7$^+$,1   & 3.938 & 16.1(10)  & 4.72(3) & 0.840 & 0.959 & 1.046 & 1.123 & 0.655 & 9.259 \\
&                             & 5$^+$,1   & 5.333 & 74.1(22)  & 4.65(1) & 0.911 & 0.982 & 1.143 & 1.338 & 0.858 &  \\
&                             & 3$^+$,1   & 5.684 & 9.6(30)  & 5.67(16) & 0.281 & 0.371 & 0.421 & 0.515 & 0.360 &  \\

$^{22}$F($4^+$) & $^{22}$Ne   & 10$^+$,2  & 3.480 & 8.7(4)  & 4.70(2) & 1.053 & 1.250 & 1.372 & 1.362 & 0.940 & 13.094 \\
&                             & 8$_1^+$,2 & 7.461 & 3.1(6) & 6.7(1)  & 0.105 & 0.106 & 0.156 & 0.099 & 0.066 &  \\
&                             & 8$_2^+$,2 & 5.500 & 53.9(6)  & 4.79(1) & 0.950 & 1.174 & 1.173 & 1.365 & 1.048 &  \\
&                             & 8$_3^+$,2 & 4.670 & 7.0(3)  & 5.34(2) & 0.504 & 0.406 & 0.107 & 0.287 & 0.706 &  \\
&                             & 8$_4^+$,2 & 3.477 & 2.45(22)  & 5.30(4) & 0.528 & 0.733 & 0.939 & 1.280 & 0.357 &  \\
&                             & 6$^+$,2   & 5.177 & 16.4(7)  & 5.26(2) & 0.553 & 0.678 & 0.649 & 0.660 & 0.382 &  \\

$^{23}$F($5/2^+$)& $^{23}$Ne  & 5$_1^+$,3 & 8.480 & 30(8)  & 5.72(16) & 0.266 & 0.377 & 0.333 & 0.321 & 0.428 & 11.953 \\
&                             & 3$_1^+$,3 & 6.660 & 10.9(19)  & 5.66(11) & 0.285 & 0.262 & 0.256 & 0.305 & 0.334 &  \\
&                             & 3$_2^+$,3 & 5.050 & 15.2(12)  & 4.96(8) & 0.637 & 0.866 & 1.028 & 1.171 & 0.881 &  \\
&                             & 3$_3^+$,3 & 4.650 & 25(4)  & 4.58(12) & 0.987 & 0.200 & 0.260 & 0.272 & 0.044 &  \\

$^{26}$F($1^+$) & $^{26}$Ne   & 4$^+$,6   & 16.170 & 36(7)  & 4.6(1)  & 0.682 & 1.145 & 1.079 & 1.065 & 0.858 & 10.691 \\
&                             & 0$^+$,6   & 18.900 & 36.5(60)  & 4.9(1)  & 0.483 & 0.733 & 0.756 & 0.714 & 0.719 &  \\

$^{23}$Ne($5/2^+$)&$^{23}$Na & 5$^+$,1   & 3.950 & 32.0(13)  & 5.38(2) & 0.393 & 0.372 & 0.392 & 0.753 & 0.656 & 9.259 \\
&                             & 3$_1^+$,1 & 4.383 & 66.9(13)  & 5.27(1) & 0.446 & 0.355 & 0.585 & 0.888 & 0.516 &  \\

$^{24}$Ne($0^+$) & $^{24}$Na  & 2$_1^+$,2 & 1.994 & 92.1(2)  & 4.35(1) & 0.525 & 0.571 & 0.442 & 0.664 & 0.053 & 4.365 \\
&                             & 2$_2^+$,2 & 1.120 & 7.9(2)  & 4.39(2) & 0.502 & 0.542 & 0.848 & 1.068 & 0.986 &  \\

$^{25}$Ne($1/2^+$)&$^{25}$Na  & 3$_1^+$,3 & 7.160 & 76.6(20)  & 4.41(2) & 0.693 & 0.751 & 0.751 & 0.697 & 0.604 & 6.901 \\
&                             & 1$_1^+$,3 & 6.180 & 19.5(20)  & 4.70(6) & 0.496 & 0.597 & 0.419 & 0.547 & 0.724 &  \\
&                             & 1$_2^+$,3 & 2.960 & 0.53(15)  & 4.82(16) & 0.432 & 0.595 & 0.713 & 1.037 & 0.657 &  \\

$^{26}$Ne($0^+$) & $^{26}$Na  & 2$_1^+$,4 & 7.258 & 91.6(2)  & 3.87(6) & 0.913 & 1.110 & 1.302 & 1.247 & 1.036 & 5.346 \\
&                             & 2$_2^+$,4 & 5.829 & 4.2(4)  & 4.8(1)  & 0.313 & 0.669 & 0.022 & 0.309 & 0.666 &  \\
&                             & 2$_3^+$,4 & 4.619 & 1.9(4)  & 4.7(1)  & 0.351 & 0.616 & 0.547 & 0.818 & 0.892 &  \\

$^{27}$Ne($3/2^+$)&$^{27}$Na  & 5$^+$,5   & 12.590 & 59.5(30)  & 4.40(4) & 0.992 & 1.229 & 1.042 & 1.059 & 1.258 & 11.548 \\

$^{28}$Ne($0^+$) & $^{28}$Na  & 2$_1^+$,6 & 12.280 & 55(5)  & 4.2(1)  & 0.624 & 1.185 & 1.104 & 1.039 & 1.147 & 6.173 \\
&                             & 2$_2^+$,6 & 10.350 & 1.7(4)  & 5.3(1)  & 0.176 & 0.816 & 0.734 & 0.352 & 0.679 &  \\
&                             & 2$_3^+$,6 & 10.160 & 20.1(12)  & 4.2(1)  & 0.624 & 0.341 & 0.484 & 0.677 & 0.333 &  \\
&                             & 2$_4^+$,6 & 9.570 & 8.5(6)  & 4.5(1)  & 0.442 & 0.471 & 0.394 & 1.293 & 0.189 &  \\

$^{24}$Na($4^+$) & $^{24}$Mg  & 8$^+$,0   & 1.392 & 99.855(5)  & 6.11(1) & 0.208 & 0.338 & 0.125 & 0.065 & 0.355 & 9.259 \\

$^{25}$Na($5/2^+$)& $^{25}$Mg & 7$^+$,1   & 2.223 & 9.48(14)   & 5.03 & 0.588 & 0.642 & 0.629 & 0.577 & 0.471 & 9.259 \\
&                             & 5$_1^+$,1 & 3.835 & 62.5(20)   & 5.26 & 0.451 & 0.558 & 0.636 & 1.169 & 0.685 &  \\
&                             & 3$_1^+$,1 & 2.860 & 27.46(22)   & 5.04 & 0.581 & 0.708 & 0.842 & 0.748 & 0.633 &  \\
&                             & 3$_2^+$,1 & 1.033 & 0.247(4)   & 5.25 & 0.457 & 0.683 & 0.712 & 1.010 & 0.910 &  \\

$^{26}$Na($3^+$) & $^{26}$Mg  & 6$_1^+$,2 & 5.413 & 1.31(4)   & 5.87(1) & 0.242 & 0.226 & 0.377 & 0.316 & 0.580 & 11.548 \\
&                             & 6$_2^+$,2 & 5.004 & 3.17(7)   & 5.33(1) & 0.450 & 0.714 & 0.536 & 1.142 & 0.504 &  \\
&                             & 6$_3^+$,2 & 3.229 & 1.72(4)   & 4.74(1) & 0.887 & 1.051 & 0.801 & 1.713 & 1.549 &  \\
&                             & 4$_1^+$,2 & 7.545 & 87.80(7)   & 4.71(1) & 0.918 & 1.073 & 0.492 & 0.865 & 0.836 &  \\
&                             & 4$_2^+$,2 & 6.416 & 0.05(4)   & 7.60(4) & 0.033 & 0.129 & 0.404 & 0.115 & 0.047 &  \\
&                             & 4$_3^+$,2 & 5.022 & 1.65(3)   & 5.62(1) & 0.322 & 0.411 & 1.037 & 0.971 & 0.552 &  \\
&                             & 4$_4^+$,2 & 4.519 & 2.738(19)   & 5.25(1) & 0.493 & 0.558 & 0.149 & 0.650 & 0.648 &  \\

\end{tabular}
\end{ruledtabular}
\end{table*}

\addtocounter{table}{-1}

\begin{table*}
\caption{{\em (Continued.)\/}}
\begin{ruledtabular}
\begin{tabular}{cccccccccccc}
 &             & &                & & & & &  M(GT)    &    \\
\cline{7-11}

$^{A}$Z$_{i}(J^{\pi})$  & $^{A}$Z$_{f}$ & 2J$_n^\pi$,2T$_n^\pi$ &  E(decay) (MeV)  & $I_\beta${(\%)}  &log$ft$(exp.) & EXPT. & USDB & IM-SRG  & CCEI   &  CEFT   & W \\
\hline

$^{27}$Na($5/2^+$)&$^{27}$Mg  & 5$_1^+$,3 & 7.310 & 11.3(7)   & 4.99(3) & 0.616 & 0.602 & 0.406 & 0.478 & 0.650 & 11.953 \\
&                             & 3$^+$,3   & 8.030 & 85.8(11)   & 4.30(15) & 1.363 & 1.747 & 1.683 & 1.562 & 1.509 &  \\

$^{28}$Na($1^+$) & $^{28}$Mg  & 4$_1^+$,4 & 12.556 & 11(6)   & 5.1(2)  & 0.384 & 0.294 & 0.438 & 0.520 & 0.245 & 9.259 \\
&                             & 2$_1^+$,4 & 9.469 & 3.2(4)   & 5.1(1)  & 0.384 & 0.536 & 0.583 & 0.592 & 0.925 &  \\
&                             & 0$_1^+$,4 & 14.030 & 60(5)   & 4.6(1)  & 0.682 & 0.840 & 0.702 & 0.773 & 0.882 &  \\
&                             & 0$_2^+$,4 & 10.168 & 20.1(19)   & 4.42(1) & 0.839 & 1.116 & 1.174 & 0.961 & 0.787 &  \\

$^{29}$Na($3/2^+$)& $^{29}$Mg & 3$^+$,5   & 13.272 & 24(8)   & 5.06(15) & 0.464 & 0.786 & 0.737 & 0.832 & 0.739 & 11.548 \\

$^{30}$Na($2^+$)  & $^{30}$Mg & 4$_1^+$,6 & 15.790 & 9.5(11)   & 5.86(6) & 0.206 & 0.408 & 0.451 & 0.509 & 0.411 & 13.803 \\

$^{27}$Mg($1/2^+$)& $^{27}$Al & 3$^+$,1   & 1.596 & 29.06(9)   & 4.934(16) & 0.381 & 0.450 & 0.373 & 0.754 & 0.468 & 5.346 \\
&                             & 1$^+$,1   & 1.766 & 70.94(9)   & 4.73(10) & 0.480 & 0.597 & 0.178 & 0.473 & 0.766 &  \\

$^{28}$Mg($0^+$) & $^{28}$Al  & 2$_1^+$,2 & 0.459 & 94.8(10)   & 4.45(9) & 0.468 & 0.624 & 0.379 & 0.819 & 0.581 & 4.365 \\
&                             & 2$_2^+$,2 & 0.211 & 4.9(10)   & 4.57(9) & 0.408 & 0.495 & 0.454 & 0.291 & 1.029 &  \\

$^{29}$Mg($3/2^+$)& $^{29}$Al & 5$_1^+$,3 & 7.613 & 27(8)   & 5.32(14) & 0.344 & 0.579 & 0.403 & 0.409 & 0.829 & 9.760 \\
&                             & 5$_2^+$,3 & 4.551 & 6.0(16)   & 4.93(13) & 0.539 & 1.064 & 1.194 & 1.769 & 1.193 &  \\
&                             & 5$_3^+$,3 & 4.428 & 28(5)   & 4.21(9) & 1.234 & 1.098 & 0.367 & 0.518 & 0.954 &  \\
&                             & 3$_1^+$,3 & 5.389 & 21(6)   & 4.73(13) & 0.678 & 0.658 & 0.341 & 0.918 & 0.187 &  \\
&                             & 3$_2^+$,3 & 4.747 & 7.8(15)   & 4.90(10) & 0.558 & 0.669 & 0.806 & 0.036 & 0.833 &  \\
&                             & 1$_1^+$,3 & 6.215 & 7(3)   & 5.49(19) & 0.283 & 0.283 & 0.198 & 0.036 & 0.067 &  \\
&                             & 1$_2^+$,3 & 4.180 & 3.0(9)   & 5.06(14) & 0.464 & 0.710 & 0.383 & 0.571 & 0.380 &  \\

$^{30}$Mg($0^+$) & $^{30}$Al  & 2$_1^+$,4 & 6.274 & 68(20)   & 3.96(13) & 0.823 & 1.203 & 1.167 & 0.994 & 1.337 & 5.346 \\
&                             & 2$_2^+$,4 & 4.549 & 7(1)   & 4.30(7) & 0.556 & 0.870 & 0.783 & 1.333 & 0.800 &  \\

$^{32}$Mg($0^+$) & $^{32}$Al  & 2$_1^+$,6 & 10.150 & 55   & 4.4  & 0.496 & 1.596 & 1.531 & 1.550 & 1.450 & 6.173 \\
&                             & 2$_2^+$,6 & 7.380 &  24.6(8)  & 4.1  & 0.700 & 0.303 & 0.385 & 0.063 & 0.190 &  \\
&                             & 2$_3^+$,6 & 6.950 & 10.7(10)   & 4.4  & 0.496 & 0.013 & 0.072 & 0.101 & 0.025 &  \\

$^{28}$Al($3^+$) & $^{28}$Si  & 4$^+$,0   & 2.863 & 99.99(1)   & 4.87(4) & 0.764 & 0.945 & 0.353 & 0.983 & 0.920 & 8.166 \\

$^{29}$Al($5/2^+$)& $^{29}$Si& 3$_1^+$,1 & 2.406 &  89.9(3)  & 5.05(5) & 0.575 & 0.924 & 0.237 & 0.388 & 0.821 & 9.259 \\
&                             & 3$_2^+$,1 & 1.253 &  6.3(2)  & 5.03(15) & 0.591 & 0.589 & 0.591 & 1.657 & 1.059 &  \\

$^{30}$Al($3^+$) & $^{30}$Si  & 6$_1^+$,2 & 3.730 & 6.6(2)   & 4.985(17) & 0.669 & 1.001 & 0.673 & 0.917 & 0.921 & 11.548 \\
&                             & 6$_2^+$,2 & 3.329 & 2.6(2)   & 5.17(4) & 0.541 & 0.657 & 0.229 & 0.869 & 0.167 &  \\
&                             & 4$_1^+$,2 & 6.326 & 17.1(9)   & 5.619(25) & 0.322 & 0.362 & 0.379 & 0.531 & 0.157 &  \\
&                             & 4$_2^+$,2 & 5.063 & 67.3(11)   & 4.578(12) & 1.069 & 1.417 & 1.192 & 0.933 & 1.189 &  \\
&                             & 4$_3^+$,2 & 3.751 & 5.7(2)   & 5.06(19) & 0.614 & 0.762 & 0.067 & 0.476 & 0.561 &  \\

$^{32}$Al($1^+$) & $^{32}$Si   & 4$_1^+$,4 & 11.080 & 4.7((13)    & 5.29(13)  & 0.308 & 0.181 & 0.122 & 0.317 & 0.414 & 9.259 \\
&                              & 4$_2^+$,4 & 8.790 &  3.0(8)   & 5.00(12)  & 0.430 & 0.590 & 0.754 & 0.449 & 0.625 &  \\
&                              & 0$_1^+$,4 & 13.020 &  85(5)   & 4.36(3)  & 0.899 & 1.093 & 0.846 & 0.952 & 1.157 &  \\
&                              & 0$_2^+$,4 & 8.040 &  4.3(11)   & 4.66(12)  & 0.637 & 1.229 & 1.336 & 1.206 & 0.938 &  \\

$^{33}$Al($5/2^+$)& $^{33}$Si  & 3$^+$,5   & 11.960 & 88(2)    & 4.3   & 1.363 & 2.225 & 1.966 & 1.491 & 2.080 & 14.143 \\

$^{31}$Si($3/2^+$)& $^{31}$P  & 1$^+$,1   & 1.491 &  99.94(7)   & 5.525(8)  & 0.272 & 0.318 & 0.076 & 0.600 & 0.389 & 7.560 \\

$^{32}$Si($0^+$) & $^{32}$P    & 2$^+$,2   & 0.227 & 100    & 8.21(6)  & 0.006 & 0.069 & 0.136 & 0.908 & 0.384 & 4.365 \\

$^{33}$Si($3/2^+$) & $^{33}$P & 1$^+$,3   & 5.845 & 93.7(7)    & 4.96(17)  & 0.520 & 0.264 & 0.394 & 0.313 & 0.329 & 9.760 \\

$^{34}$Si($0^+$) & $^{34}$P    & 2$^+$,4   & 2.984 & 100    & $\geq$3.3   & 1.759 & 0.639 & 0.372 & 0.845 & 0.232 & 5.346 \\

$^{32}$P($1^+$) & $^{32}$S     & 0$^+$,0   & 1.710 & 100    & 7.90(2)  & 0.015 & 0.038 & 0.140 & 0.543 & 0.352 & 5.346 \\

$^{33}$P($1/2^+$) & $^{33}$S   & 3$^+$,1   & 0.248 & 100    & 5.022(7)  & 0.343 & 0.295 & 0.025 & 0.336 & 0.337 & 5.346 \\

$^{34}$P($1^+$) & $^{34}$S     & 4$_1^+$,2 & 3.255 & 14.8(20)    & 4.93(6)  & 0.467 & 0.742 & 0.533 & 0.593 & 0.837 & 7.560 \\
&                              & 4$_2^+$,2 & 1.268 & 0.31(6)    & 4.88(9)  & 0.494 & 0.082 & 0.313 & 1.573 & 0.229 &  \\
&                              & 0$_1^+$,2 & 5.383 & 84.8(21)    & 5.159(12) & 0.358 & 0.480 & 0.180 & 1.545 & 0.269 &  \\

\end{tabular}
\end{ruledtabular}
\end{table*}

\begin{table}
\caption{\label{qf_tab}Quenching factor for the different effective interactions. }
\begin{ruledtabular}
\begin{tabular}{ccc}
Interaction   &  $q$  & RMS deviations \\
\hline
USDB   & 0.77$\pm$0.02   & 0.0356 \\
IM-SRG  & 0.75$\pm$0.03   & 0.0469\\
CCEI   & 0.62$\pm$0.03   & 0.0440\\
CEFT   & 0.73$\pm$0.04   & 0.0541\\
\end{tabular}
\end{ruledtabular}
\end{table}

\begin{figure*}
\includegraphics[width=8.0cm,height=6.2cm,clip]{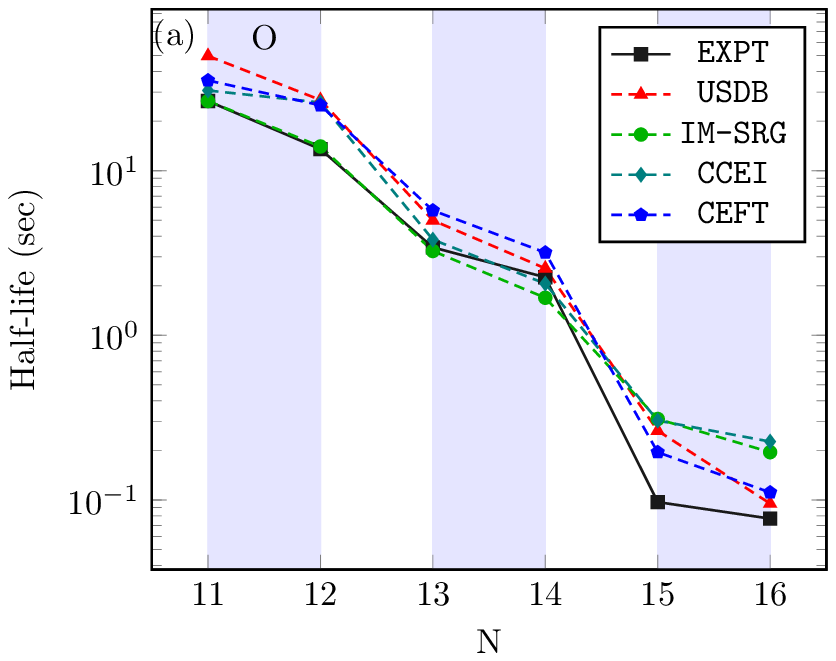} 
\includegraphics[width=8.0cm,height=6.2cm,clip]{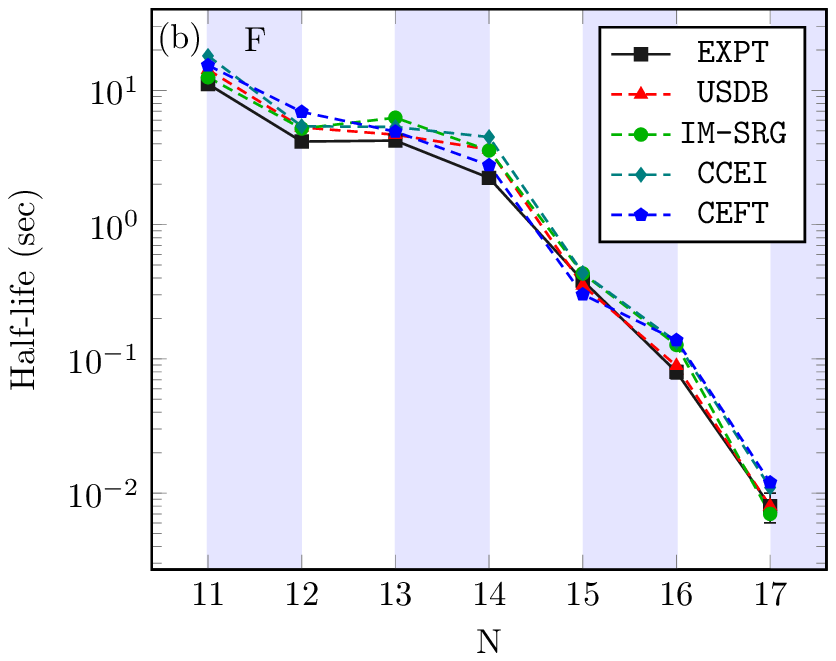} 
\includegraphics[width=8.0cm,height=6.2cm,clip]{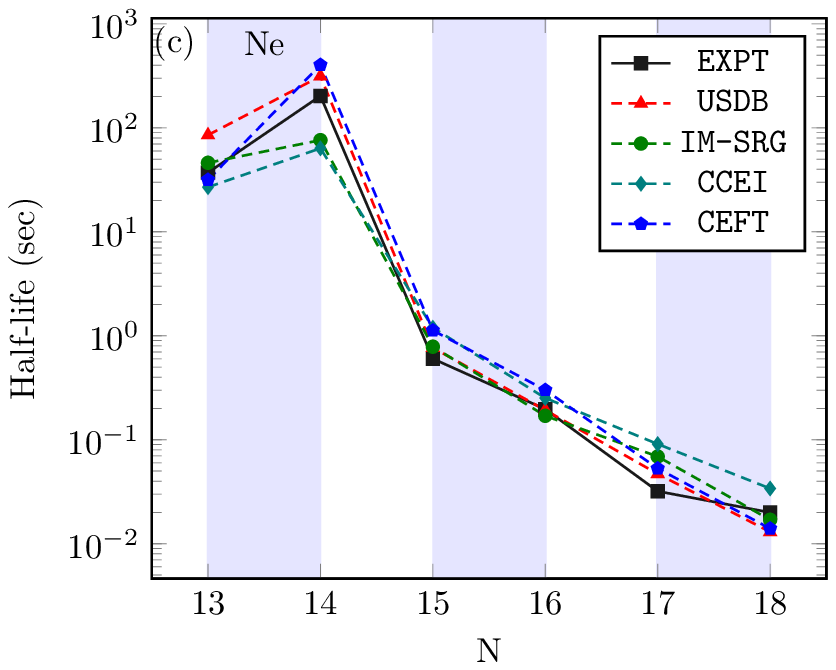} 
\includegraphics[width=8.0cm,height=6.2cm,clip]{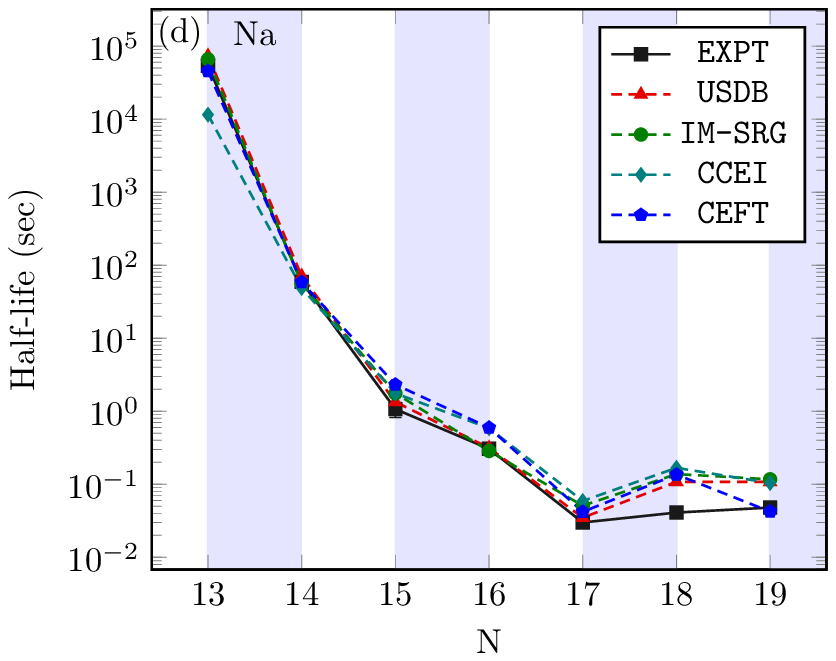}
\includegraphics[width=8.0cm,height=6.2cm,clip]{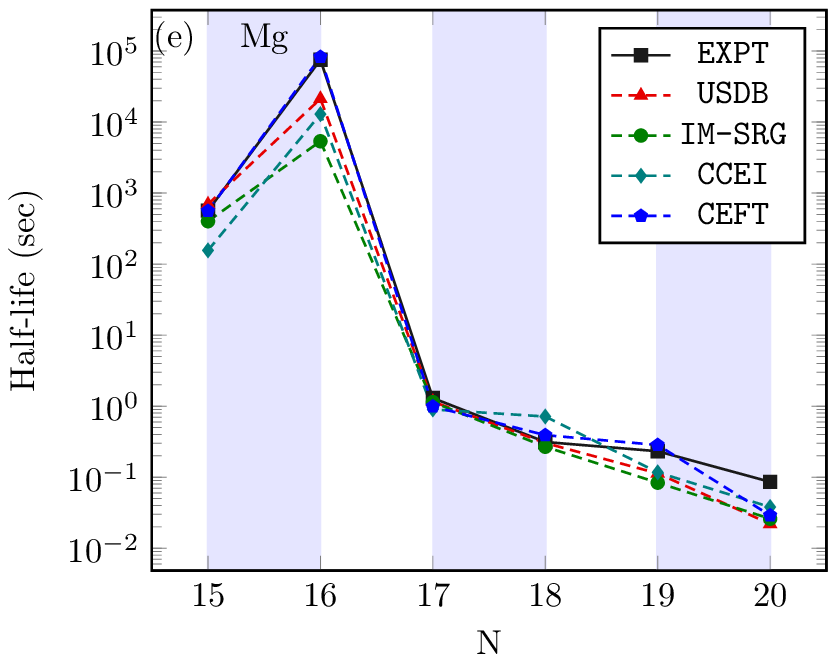}
\includegraphics[width=8.0cm,height=6.2cm,clip]{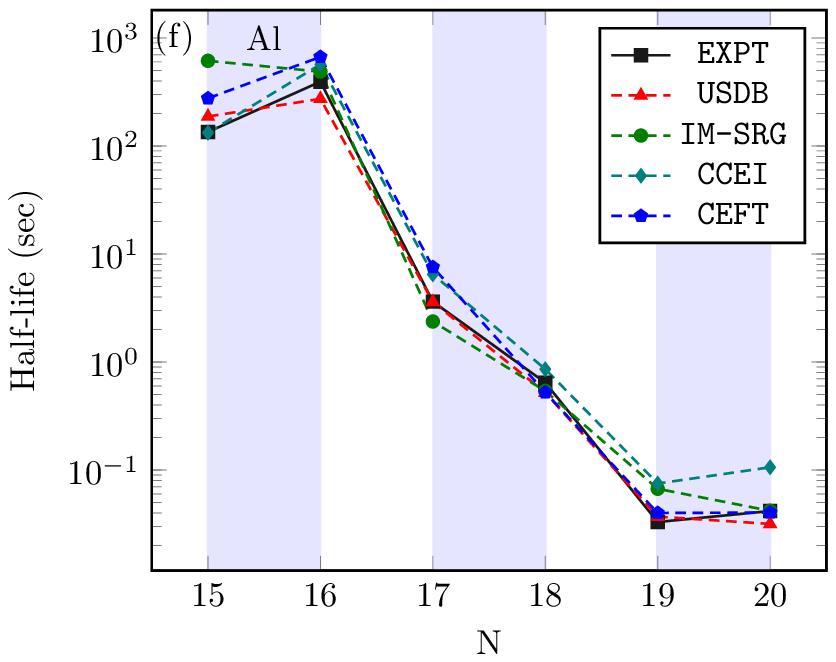} 
\includegraphics[width=8.0cm,height=6.2cm,clip]{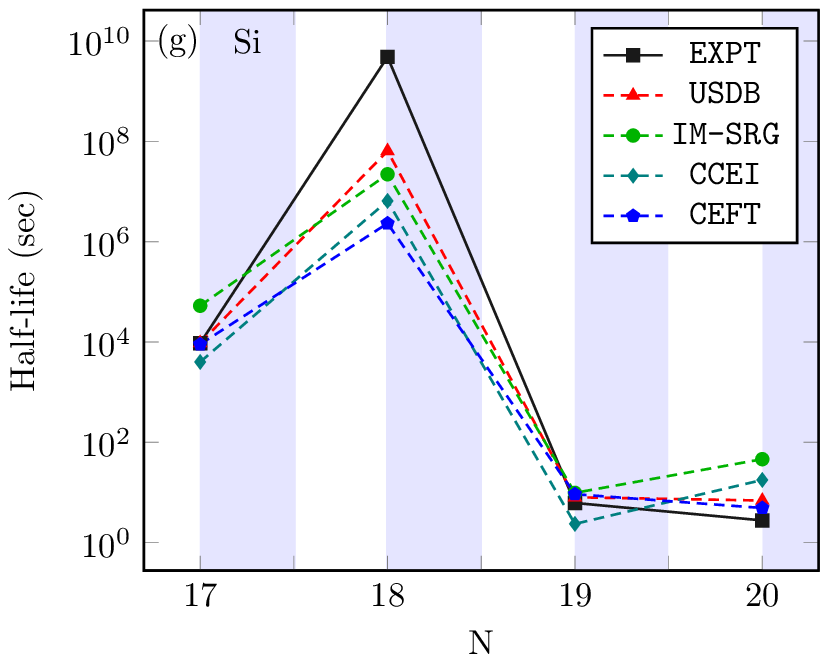} 
\includegraphics[width=8.0cm,height=6.2cm,clip]{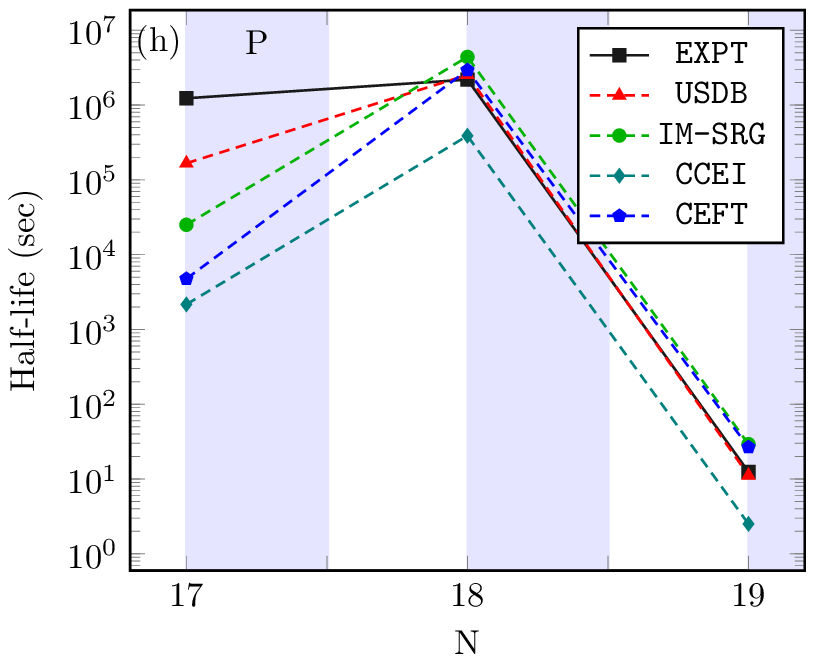} 
\caption{\label{hf} (Color online) 
Plot of half-lives  versus neutron numbers of $sd$ shell nuclei.}
\end{figure*}

In Fig.~ \ref{logft}   we show the distribution of calculated log$ft$ values with the experimental data for
some $\beta^{-}$--decays nuclei for which experimental log$ft$ values are available for excited states also.
 In case of $^{21}$F, although  results of the $ab~initio$ interactions for excitation energy for the excited $3/2^+$,
$5/2^+$ and $7/2^+$ states slightly differ from the experimental data, all the 
interactions give calculated log$ft$ values close to the experimental data.
For $^{28}$Ne, the calculated log$ft$ values with  the CCEI  are better in comparison to other interactions,
 although CCEI and IM-SRG interactions are not able to reproduce correctly energy levels in comparison to USDB and CEFT interactions. 
The calculated value for excitation energy for $2_1^+$ state 
is in good agreement  for all the interactions for $^{28}$Na. 
For this nucleus all the four interactions  give reasonable results for log$ft$ values.

In Fig. \ref{hf}, we compare the theoretical and  the experimental $\beta$-decay half-lives of 
 $sd$ shell  nuclei. 
 Here, we make some general comments on the half-lives.
(1) For O and F isotopes, calculated half-lives are in fair agreement with the experimental values within a factor of 2.1-2.2,
except for $^{22}$O obtained with IM-SRG.
(2) The discrepancy between calculated and experimental half-life becomes large (a) when the discrepancy between the calculated and experimental $B(GT)$ is large, or 
(b) when the transition with the dominant branching  ratio is different between the calculation and the experiment, 
or (c) when the $Q$ value for the transition is small and the difference between the calculated and experimental 
excitation energies is large enough to lead to a substantial change of the phase space  factor for the transition.
In case of $^{22}$O with IM-SRG, a large discrepancy comes from combined 
effects of (a) and (b).    
Nuclei in the island of inversion such as $^{32}$Mg can not be well described for both the $ab~initio$ and phenomenological interactions due to the reason (a).
$^{28}$Mg discussed above corresponds to the case (c). 
 (3) For isotopes with $Z$ =10-13 ($Z$=14-15), there are one or two cases (or more cases) for each $Z$ in
 which the calculated half-lives differ from the experimental ones by a factor more than 3 due to the reasons (a), (b) or (c)  in case of IM-SRG and CCEI. 
  In general calculated half-lives results for O, F, Ne, Na, Mg and Al
are in reasonable agreement with the experimental data.  
The results of P isotopes are showing deviation with the experimental data. This might be due to missing of the $pf$ orbitals in our calculations. 

The phase space  factor, which is estimated to be roughly proportional to ($Q$)$^5$. 
For example in the case of 
 $^{28}$Mg(0$^+$) $\rightarrow$ $^{28}$Al($1^+$) transition, the excitation energies for the 1$_{1}^{+}$ state of $^{28}$Al obtained for the 
 interactions are smaller than the experimental one,
$E_x$ =1.373 MeV, by 0.175, 0.571, 0.251 and 0.018 MeV for USDB, IM-SRG, CCEI and CEFT, respectively, which leads to an enhancement of 
the phase space  factor by nearly 10 times for IM-SRG.   
Though the difference of the B(GT) values is within a range of a factor of about 3, 
large difference in the phase space  factors leads to larger difference in the half-lives.

\section{Summary and conclusions}

 In the present work we have performed shell model calculations using $ab~initio$ approaches along with interaction based on 
chiral effective field theory and phenomenological USDB interaction, and evaluated $B(GT)$, 
log$ft$ values and  half-lives
for the $sd$ shell nuclei. 
Since these 
 $ab~initio$ effective interactions are developed using  state-of-the-art approaches,
 our aim is to test the predicting power of these interactions for $\beta^-$- decay properties.

 We find that all the $ab~initio$ interactions  as well as the  fitted interaction based on chiral effective theory considered here need certain quenching of the $GT$ strengths, as large as by 44-62$\%$, as for the phenomenological USDB interaction.

 The quenching factor can be attributed to (i) configurations outside the $sd$ shell, 
 (ii) induced effective Gamow-Teller operator due to the unitary transformation in the $ab~initio$ approach, and
 (iii) the intrinsic two-body Gamow-Teller operator connected with 3N interaction.
  
 An effective value needs to be used for the weak axial coupling constant $g_{A}$ in the shell model calculation.
In our calculation corresponding to $|g_A^{\text{free}}|=1.26$,  we have obtained the value of
$|g_A^{\text{eff}}|$ by chi-squared fitting as  0.95,  0.78, and 0.92 for IM-SRG,  CCEI, and CEFT interactions, respectively,
while it has been obtained as 0.97 for the USDB.
 We have also obtained the  RMS deviation values 0.0469, 0.0440, 0.0541, and 0.0356 corresponding to IM-SRG, CCEI, CEFT and USDB interactions, respectively. 
The RMS deviations for the $ab~initio$ and CEFT interactions are slightly enhanced compared with the USDB interaction.
Within the present framework of using the one-body operator with effective $g_A$, the $ab~initio$ IM-SRG interaction can be considered as most favorable since both the quenching
factor and the RMS deviation are close to those of the phenomenological USDB. 

The calculated half-lives results for O, F, Ne, Na, Mg and Al
are in reasonable agreement with the experimental data  for the $ab~initio$ interactions. 
 However, it is rather hard to reproduce both the transition strength and energy levels in nuclei such as $^{28}$Ne and $^{32}$Mg in the island of inversion.

 In case of $ab~initio$ approaches, IM-SRG and CCEI, both the intrinsic and induced two-body operators can be constructed in principle, 
and the contributions from configurations outside the $sd$-shell can be reliably constrained. The quenching factors can become closer to unity with the inclusion of the two-body operators \cite{Gysbers}.
It would be also interesting to expand the configuration space outside the $sd$-shell so that we can treat nuclei in the island of inversion more appropriately.






\section*{Acknowledgments}
AK acknowledges financial support from MHRD for his Ph.D. thesis work.
PCS would like to thank Prof. Gabriel Mart\'inez Pinedo and Prof. Thomas Neff for useful discussion on this work. 
TS would like to thank a support from  JSPS KAKENHI under Grant No. 15K05090.


\end{document}